\documentclass[journal, onecolumn, 11pt, draftcls]{IEEEtran}

\usepackage{ifthen}
\usepackage[cmex10]{amsmath}
\usepackage{amsthm,amssymb}
\usepackage{epsfig}

\newtheorem{theorem}{Theorem}
\newtheorem{lemma}{Lemma}
\newtheorem{proposition}{Proposition}

\begin{document}

\title{Optimized data sharing in multicell MIMO with finite backhaul capacity}

\author{Randa Zakhour and David Gesbert \thanks{Randa Zakhour is with the ECE Department of the National University of Singapore (NUS). David Gesbert is with the Mobile Communications Department of EURECOM, France. The research leading to these results has received funding from the European Commission's seventh framework programme FP7-ICT-2009 under grant agreement $\textrm{n}^\circ$ 247223 also referred to as ARTIST4G. The work was also partly funded from NUS grant R-263-000-572-133, Singapore.
}
}

\maketitle

\begin{abstract}
This paper addresses cooperation in a multicell environment where base stations (BSs) wish to jointly serve multiple users, under a constrained-capacity backhaul. We point out that for finite backhaul capacity a trade-off between sharing user data, which allows for full MIMO cooperation, and not doing so, which reduces the setup to an interference channel but also requires less overhead, emerges. We optimize this trade-off by formulating a rate splitting approach in which non-shared data (private to each transmitter) and shared data are superposed. We derive the corresponding achievable rate region and obtain the optimal beamforming design for both shared and private symbols. We show how the capacity of the backhaul can be used to determine how much of the user data is worth sharing across multiple BSs, particularly depending on how strong the interference is.
\end{abstract}

\section{Introduction}
Interference is a major issue in several types of wireless networks. The related problem is especially acute in cellular networks with full spectrum reuse across all base stations (BSs) (see \cite{gesbert_jsac10} and references therein). In traditional designs, each BS obtains from the backhaul the data intended for users in its coverage area alone, i.e.  if one ignores cases of soft handover, data for users is not available at multiple BSs: this results in the so-called interference channel (IC) and was treated for the MISO case in \cite{jorswieck_sp08} and \cite{dahrouj_tw10} for example.  Recent research rooted in MIMO theory has suggested the benefits of relaxing this constraint, allowing for user messages to be shared at multiple transmitters so that a giant broadcast MIMO channel ensues. In such a scenario, multicell processing in the form of joint precoding is realized, e.g. for the downlink: this scheme is referred to as network MIMO (a.k.a. multicell MIMO) (see \cite{shamai_vtc2001}, \cite{karakayali_icc06}, \cite{somekh_globecom06} for example).

In this paper, we focus on the issue of data sharing in a multicell cooperation setup. Other issues, such as the complexity of a centralized implementation of network MIMO or CSI sharing are for example tackled in \cite{ng_it2008} and \cite{bjornson_sp2010}, respectively.

In fact, full data sharing subsumes high capacity backhaul links, which may not always be available, or even simply desirable. In fact, under limited backhaul rate constraints, data sharing consumes a precious fraction of the backhaul capacity which otherwise could be used to carry more data to the users: this overhead should thus be compensated by the capacity gain induced by the network MIMO channel over the classical IC.

A number of recent interesting research efforts have considered networks with finite-capacity backhaul. To cite a few, in \cite{shamai_itaw08} and \cite{simeone_eurasip09}, joint encoding for the downlink of a cellular system is studied under the assumption that the BSs are connected to a central unit via finite-capacity links. The authors investigate different transmission schemes and ways of using the backhaul capacity  in the context of a modified version of Wyner's channel model. One of their main conclusions is that "central encoding with oblivious cells", whereby quantized versions of the signals to be transmitted from each BS, computed at the central unit, are sent over the backhaul links, is shown to be a very attractive option for both ease of implementation and performance, unless high data rate are required. If this is the case, the BSs need to be involved in the encoding, i.e. at least part of the backhaul link should be used for sending the messages themselves not the corresponding codewords.

In \cite{marsch_ew07}, an optimization framework, for an adopted backhaul usage scheme, is proposed for the downlink of a large cellular system. A so-called joint transmission configuration matrix is defined: this specifies which antennas in the system serve which group of users. The backhaul to each BS is used to either carry quantized versions of the transmit signals computed centrally similarly to the scheme in \cite{simeone_eurasip09}, except that a more realistic system model is assumed; alternatively, the backhaul is used to carry uncoded binary user data.

In \cite{marsch_globecom08}, a more information-theoretic approach is taken and a two-cell setup is considered in which, in addition to links between the network and each BS, the two multi-antenna BSs may be connected via a finite-capacity link:  different usages of the backhaul are optimized and their rate regions compared under suboptimal maximum ratio transmission beamforming.
\cite{marsch_globecom09} uses duality theory to optimize transmission for both quantized and unquantized message based cooperation schemes: for the unquantized case, intermediate schemes that time-share between no and full cooperation while meeting the backhaul constraints are proposed.

Imposing finite capacity constraints on the backhaul links brings with it a set of interesting research as well as practical questions, since more cooperation between BSs is expected in 4G cellular networks, in particular:
\begin{itemize}
\item Given the backhaul constraints, assuming that {\em not all} traffic is shared across transmitters, i.e assuming a certain part remains private to each transmitter,  what rates can be achieved?
\item How useful is data sharing when backhaul constraints are present? I.e., how do the rates achieved with a data sharing joint transmission enabling scheme compare to those achieved without data sharing, under limited backhaul?
\item Is there a backhaul capacity lower bound below which it does not pay off to share user data?
\end{itemize}

In this work, we attempt to answer these questions by considering a  setup in which a finite rate backhaul connects the network with each of the BSs, and focusing on how to use this given backhaul to serve the users in the system. To simplify exposition, we focus on the two-cell problem. A transmission scheme is specified whereby superposition coding is used to transmit signals to each user:
each user's data is in fact split into two types, `private' data sent by a single BS and  `shared' data transmitted via multiple bases.
Thus for all non trivial traffic ratios between private and shared data, our system corresponds to a hybrid channel, which in an information theoretic sense may be considered as an intermediate between the MIMO broadcast (or "network MIMO") and the interference channels.
 The corresponding rate region is expressed in terms of the backhaul constraints and the beamforming vectors used to carry the different signals: finding the boundary of the aforementioned region is reduced to solving a set of convex optimization problems. In doing so, we also solve the problem of optimal beamforming design for this hybrid IC/MIMO channel. We compare the rates achieved in such a rate splitting scheme to those obtained for network MIMO and the IC and illustrate the gains related to moderate sharing levels in certain realistic situations.  An alternative approach to the use made here of the backhaul is the one proposed in \cite{simeone_eurasip09} for example: as described earlier, such an approach assumes the BSs are oblivious to the encoding and uses the backhaul to forward quantized versions of the encoded symbols to be transmitted over the air. We also adapt this scheme to our setup and compare its performance to our rate splitting scheme.

\section{System Model and Proposed Transmission scheme}
\begin{figure}[htb]
\center{
\includegraphics[width=3.5in, height=2.7in, viewport = 15 70 700 500, clip]{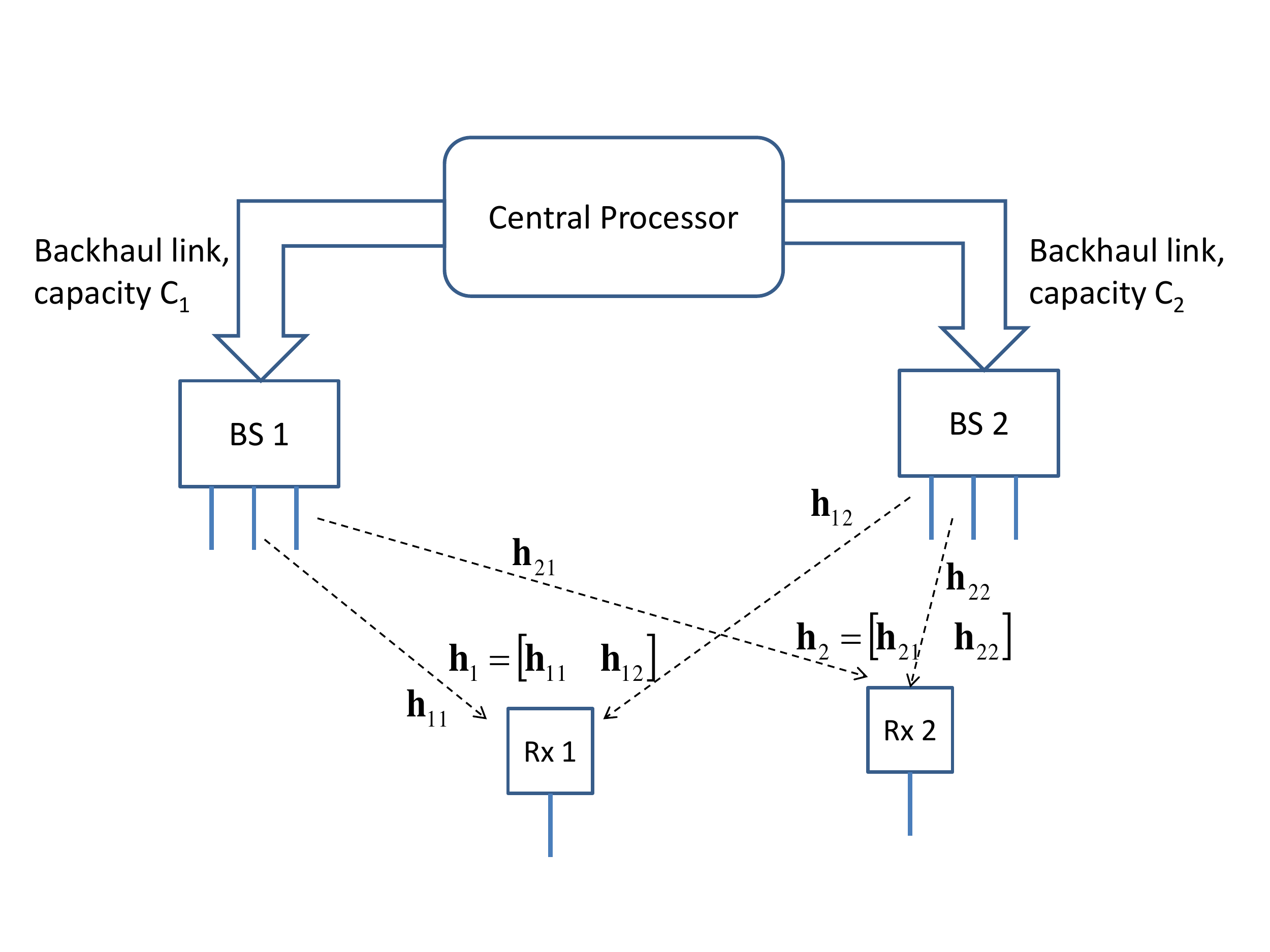}
\caption{Constrained backhaul setup. The rates of the messages carried by each backhaul link are represented. The central processor is assumed to collect all downlink traffic then routes it to individual (non shared traffic) or both (shared traffic) transmitters.} \label{fig:scenario}
}
\end{figure}

The system considered is shown in Figure \ref{fig:scenario}. In this study, we focus on a two transmitter two receiver setup.
As we emphasize the problem of precoding at the transmitter side, the receivers are assumed to have a single antenna whereas transmitters have $N_t \ge 1$ antennas each: $\mathbf{h}_{ij}$ is the $N_t$-dimensional complex row vector corresponding to the channel between transmitter $j$ and user $i$; $\mathbf{h}_i = \left[\mathbf{h}_{i1}~ \mathbf{h}_{i2}\right]$ represents user $i$'s whole channel state vector.
The signal received at user $i$ will be given by
\begin{align}
y_i = \sum_{j=1}^2\mathbf{h}_{ij} \mathbf{x}_j + z_i, \label{eq:y_i}
\end{align}
where $\mathbf{x}_j \in \mathbb{C}^{N_t}$ denotes BS $j$'s transmit signal, and $z_i \sim \mathcal{CN}(0, \sigma^2)$ is the receiver noise. $\mathbf{x}_j$ is subject to power constraint $P_j$ so that
\begin{align}
\mathbb{E} \|\mathbf{x}_j\|^2 \le P_j, \quad j= 1,2. \label{eq:PConstraint}
\end{align}

We assume a backhaul link of capacity $C_j$ [bits/sec/Hz] between the central processor (CP)  or the backbone network, and transmitter $j$, for $j = 1, 2$: the central processor collects all downlink traffic then routes it to individual (non shared traffic) or both (shared traffic) transmitters.
In an attempt to bridge the interference channel situation (where the transmitters do not share user data) and the multi-cell MIMO scenario (where they do), we propose to split the user traffic content across two types of messages:
\begin{itemize}
\item \emph{private messages} will be sent from the CP to only one of the transmitters, and
\item \emph{shared messages}, which are sent from the CP to both transmitters, and are consequently jointly transmitted.
\end{itemize}
Note that while previous work (e.g. \cite{marsch_globecom08}) has considered private and common messages in a similar sense to that in the Han-Kobayashi approach to the interference channel, i.e. where common messages are transmitted by a single BS but decoded by both receivers, here the private messages and shared messages are fundamentally different and are instead analogous to the private and common messages in Slepian-Wolf's Multiple Access Channel (MAC) with a common message \cite{slepian_Bell73}, where a common message is meant for a single receiver but known at both transmitters.

Thus, the total information rate for user $i$, $r_i$, will be split across $r_{i1, p}$, $r_{i2, p}$ and $r_{i, c}$, where $r_{i, c}$ refers to the rate of the shared message for that user, and $r_{ij, p}$ refer to the rate of the private message for user $i$ reaching it from BS $j$:
\begin{align}
r_i = \sum_{j=1}^2 r_{ij,p} + r_{i,c}. \label{eq:Ri}
\end{align}

\emph{Assumptions} We assume each receiver does single user detection (SUD), in the sense that
any source of interference is treated as noise. Note that this paper examines the costs and benefits of sharing user data, not that of sharing the channel state information  (CSI), hence full global  CSIT is assumed at each transmitter.
 A companion paper which focuses on the problem of CSIT sharing can be found in \cite{zakhour_itaw10}.

\subsection{Particular Cases}
The transmission scheme introduced here covers the two particular cases of:
\begin{itemize}
\item An interference channel, obtained by forcing $r_{ii,p} \equiv r_{i}, i = 1, 2$, and
\item A network MIMO channel, obtained by forcing $r_{ij,p} \equiv 0, i = 1, 2, j = 1,2$.
\end{itemize}

\emph{Notation} In what follows,  $\bar{i} = \mod(i, 2)+1, i = 1, 2$ and is used to denote the \emph{other} transmitter/receiver depending on the context.

\subsection{Backhaul usage}
Here we introduce some fundamental inequalities imposed by the backhaul constraints which will be helpful in characterizing the achievable rate region for this hybrid IC/MIMO channel.
Backhaul link $j$ with finite capacity $C_j$ serves to carry both private (from BS $j$) and shared messages for both users, so that the following constraint applies:
\begin{align}
C_j \ge \sum_{i=1}^2 r_{ij,p} + \sum_{i=1}^2 r_{i,c}, \quad j = 1, 2.
\end{align}

Using \eqref{eq:Ri}, this constraint can be rewritten as:
\begin{align}
C_j \ge \sum_{i=1}^2 r_{i}-  \sum_{i=1}^2 r_{i\bar{j}, p}, \quad j = 1, 2. \label{eq:Ci}
\end{align}
Finally, the sum rate $r = r_1 + r_2$, cannot exceed the total backhaul capacity, so that
\begin{align}
r \le C_1 + C_2.
\end{align}

\subsection{Over-the-air transmission}
The channel between the two transmitters and user $i$ can be viewed as a MAC with a common
message \cite{slepian_Bell73}. The overall channel can be regarded as the superposition of two such channels, which interfere with each other so that the receiver noise at user $i$ is enhanced by the interference due to the signals carrying  user $\bar{i}$'s data; the total interference plus noise power at user $i$ will be denoted by $\sigma^2_i$.
We thus write the transmit signal of BS $j$ as a superposition of two signals, $\mathbf{x}_{ij}$, $i=1,2$, one intended for each user:
\begin{align}
\mathbf{x}_j = \sum_{i=1}^2 \mathbf{x}_{ij}.
\end{align}

Restricting the transmission model to beamforming, $\mathbf{x}_{ij}$ can be generated as:
\begin{align}
\mathbf{x}_{ij} &= \mathbf{w}_{ij,c} s_{i,c} + \mathbf{w}_{ij,p} s_{ij,p}, \label{eq:xij}
\end{align}
where $s_{i,c}$ and $s_{ij,p}$ are independent $\mathcal{CN}(0,1)$ random variables; $\mathbf{w}_{i,c} = [\mathbf{w}_{i1,c}^T \mathbf{w}_{i2,c}^T]^T \in \mathbb{C}^{2N_t}, $ is the beamforming vector carrying symbols $s_{i,c}$, and $\mathbf{w}_{ij,p} \in \mathbb{C}^{N_t}$ is the beamforming vector carrying symbols $s_{ij,p}$.

The following proposition specifies a rate region corresponding to the over-the-air channel, achievable by transmit signals of the form given in \eqref{eq:xij}. Details of the proof can be found  in Appendix \ref{app:mac_common}.

\begin{proposition}\label{prop:oA} The following rate region $\mathcal{R}_{\textrm{air}}$ is achievable on the over-the-air segment:
\begin{align}
&r_{ij,p} \le \log_2 \left(1 + \frac{\left|\mathbf{h}_{ij}\mathbf{w}_{ij,p}\right|^2}{\sigma^2_i}\right), \quad j = 1,2, i = 1,2 \nonumber \\
&\sum_{j=1}^2 r_{ij,p} \le \log_2 \left(1 + \frac{\sum_{j=1}^2 \left|\mathbf{h}_{ij}\mathbf{w}_{ij,p}\right|^2}{\sigma^2_i}\right), \quad i = 1, 2\nonumber \\
&r_{i} \le \log_2 \left(1 + \frac{\left|\mathbf{h}_{i}\mathbf{w}_{i,c}\right|^2+\sum_{j=1}^2 \left|\mathbf{h}_{ij}\mathbf{w}_{ij,p}\right|^2}{\sigma^2_i}\right), \quad i = 1, 2
\label{eq:MAC_rates}
\end{align}
where 
\begin{align}
\sigma^2_i 
&=  \sigma^2 + \sum_{j=1}^2 \left|\mathbf{h}_{ij}\mathbf{w}_{\bar{i}j,p}\right|^2 +\left|\mathbf{h}_{i}\mathbf{w}_{\bar{i},c}\right|^2, \label{eq:sigmai2}
\end{align}
and the beamforming vectors are subject to power constraint 
\begin{align}
\sum_{i=1}^2 \left(\|\mathbf{w}_{ij,c}\|^2 + \|\mathbf{w}_{ij,p}\|^2\right) \le P_j, \quad j = 1, 2. \label{eq:powerCon2}
\end{align}
\end{proposition}

\section{Rate Region}
An achievable rate region $\mathcal{R}$ is the set of $(r_1, r_{11,p}, r_{12,p}, r_2, r_{21,p}, r_{22,p})$, as specified above, that satisfy the specified backhaul and power constraints. We are particularly interested in the boundary of the rate region, in maximum weighted sum rate points, and in the beamforming strategies to reach these points.
As the direct characterization of the rate region is a difficult task here, one may obtain the rate region boundary by using the rate profile notion from \cite{mohseni_jsac06}: a rate profile specifies how the total rate is split between the users. Points on the rate region boundary are thus obtained by solving the following problem for $\alpha$ discretized over $[0, 1]$, where $\alpha$ denotes the proportion of the total sum rate intended for user 1's data:

\begin{align}
\text{max. } &r \nonumber \\
\text{s.t. } &r_1 = \alpha r,  \quad r_2 = (1-\alpha) r \nonumber \\
& r_i \ge 0, r_{ij,p} \ge 0, i = 1,2, j = 1, 2 \nonumber \\
& \sum_{j=1}^2 r_{ij,p} \le r_i, i = 1, 2 \nonumber \\
& \sum_{i=1}^2 r_{i}-  \sum_{i=1}^2 r_{i\bar{j}, p} \le  C_j, \quad j = 1,2 \nonumber \\
&\left(r_1, r_{11,p}, r_{12,p}, r_2, r_{21,p}, r_{22, p}\right) \in \mathcal{R}_{\textrm{air}},
\end{align}
where $\mathcal{R}_{\textrm{air}}$ was defined in Proposition \ref{prop:oA} and the remaining constraints follow from \eqref{eq:Ri} and \eqref{eq:Ci}.

This problem may be solved using a bisection method over $r$, which requires testing the feasibility of any chosen sum rate $r$: the latter is detailed in subsection \ref{sec:FeasibleSR} below.

\subsection{Establishing feasibility of a given rate pair $(r_1, r_2)$} \label{sec:FeasibleSR}
Assume sum rate $r$ and $\alpha$ to be fixed. Thus, $r_1 = \alpha r$, $r_2 = (1-\alpha) r$.  An important question toward characterizing the rate region boundary is how do we establish feasibility of this rate pair?

\begin{lemma}\label{lemma1}
Rate pair $(r_1, r_2)$ is achievable, if a rate-tuple $(r_1, r_{11,p}, r_{12,p}, r_2, r_{21,p}, r_{22,p})$ such that
\begin{align}
\sum_{j=1}^2 r_{j\bar{i},p} = \max\left(0, r_1+r_2-C_{i}\right) \equiv c_{\bar{i}}, \label{eq:Rmin}
\end{align}
can be supported on the over-the-air segment.
\end{lemma}
\begin{IEEEproof}
On the over-the-air segment of our communication setup, more data sharing increases the achievable rate region. On the other hand, the backhaul constraints limit the amount of shared data. From Equation \eqref{eq:Ci}, the private rates must satisfy the constraint that $\sum_{j=1}^2 r_{j\bar{i},p}$ be no less than $c_{\bar{i}}$.
\end{IEEEproof}

\begin{lemma}\label{lemma:powMin}
Feasibility of a rate pair $(r_1, r_2)$ may be checked by solving the following power minimization: 
\begin{align}
\text{min. } & \sum_{i=1}^2 \sum_{j=1}^2 \left(\|\mathbf{w}_{ij,c}\|^2 + \|\mathbf{w}_{ij,p}\|^2\right) \label{eq:minPow} \\
\text{s.t. } & 0 \le r_{1j,p} \le c_j, ~ j = 1,2 \nonumber \\
& c_1 + c_2 - r_2 \le r_{11,p} + r_{12,p} \le r_1 \nonumber \\
&\left(r_1, r_{11,p}, r_{12,p}, r_2, c_1-r_{11,p}, c_2-r_{12,p}\right) \in \mathcal{R}_{\textrm{air}}.\nonumber
\end{align}
\end{lemma}
\begin{IEEEproof}
Problem \eqref{eq:minPow} is a power minimization subject to power and rate constraints $(r_1, r_2)$ taking into consideration Lemma \ref{lemma1}. The latter gives
\begin{align}
r_{11,p} = c_1 - r_{21,p}, \nonumber \\
r_{12,p} = c_2 - r_{22,p}.
\end{align}
Combining this with the constraints
\begin{align}
r_i \ge \sum_{j=1}^2 r_{ij,p}, i = 1, 2, \label{eq:RiConstraint}
\end{align}
and the nonnegativity constraints of all rates, we obtain the following constraints on $r_{11,p}$ and $r_{12,p}$:
\begin{align}
0 \le r_{11,p} \le c_1 \nonumber\\
0 \le r_{12,p} \le c_2 \nonumber \\
c_1 + c_2 - r_2 \le r_{11,p} + r_{12,p} \le r_1. \label{eq:polyhedron}
\end{align}
\end{IEEEproof}

\noindent \emph{Solving Problem \eqref{eq:minPow}}\\
From Lemma \ref{lemma:powMin}, we can establish feasibility of rate pair $(r_1, r_2)$ by solving \eqref{eq:minPow}. Fixing all rates in \eqref{eq:minPow}, the remaining power minimization problem can be shown to be equivalent to a convex optimization, and thus solved efficiently. This is discussed in more detail in Appendix \ref{app:Dual}.

Constraints \eqref{eq:polyhedron} define a polyhedron. If a single $c_i$ is zero, $r_{1i,p} = 0$, and the polyhedron collapses down to a line segment. If both $c_i$'s are zero, the line segment further collapses into a single point: in this case, there are no private messages and only shared messages are needed, as in the traditional network MIMO setup, and only the $(r_{11,p}, r_{12,p})$ pair $(0, 0)$ needs to be checked for feasibility.
In the general case,  we conjecture that to solve \eqref{eq:minPow}, it is enough to check feasibility at the corner points of the polytope defined in \eqref{eq:polyhedron}, so that to check feasibility we only need to solve the above convex optimization at a small number of points, which is what we do in our simulations.

Note that the above rate region could further be expanded by dirty-paper coding \cite{caire_it2003} the shared messages, thereby reducing the interference at one of the users: thus,
if user 1's shared message is encoded first, dirty-paper coding of the other messages at user 2 ensures the corresponding signal does not cause interference to user 2.

\section{Quantized Backhaul (Oblivious Base Stations)}	
Depending on the network setup, it may also be possible to move the processing away from the BSs and assume these to be ignorant of the encoding scheme: this type of framework was recently proposed in \cite{simeone_eurasip09}, where the dirty paper encoded then quantized input signal in a Wyner-type channel model network is optimized. In this section, we modify the scheme in \cite{simeone_eurasip09}
 to linear beamforming in our channel setup. In this oblivious BSs configuration, the CP designs the transmission and the backhaul is now used to carry to each transmitter a quantized version of the signals it should transmit.

\subsection{Oblivious BS with Linear Beamforming}
Before any quantization takes place and for a linear precoding scheme, the signal to be transmitted by base station $i$, $\mathbf{x}_i \in \mathbb{C}^{N_t}$ can be written as
\begin{align}
\mathbf{x}_i = \mathbf{w}_{1i} s_i + \mathbf{w}_{2i} s_2,
\end{align}
where $s_k \sim \mathcal{CN}(0,1)$ are the symbols carrying user $k$'s message. Thus,
\begin{align}
\mathbf{x}_j  \sim \mathcal{CN}\left(\mathbf{0}_{N_t}, \mathbf{w}_{1j}\mathbf{w}_{1j}^H + \mathbf{w}_{2j}\mathbf{w}_{2j}^H\right),
\end{align}

Since the backhaul links have finite capacity, the designed $\mathbf{x}_j$ may not be forwarded perfectly to the corresponding BS. Thus, quantization is resorted to. This may be modeled, as in \cite{simeone_eurasip09}, by a forward test channel of the form
\begin{align}
\hat{\mathbf{x}}_j = \mathbf{x}_j + \mathbf{q}_j, \label{eq:xQ}
\end{align}
where $\mathbf{x}_j$ is as specified above and $\mathbf{q}_j$ is the quantization noise, independent of $\mathbf{x}_j$ and such that $\mathbf{q}_j \sim \mathcal{CN}\left(\mathbf{0}, \mathbf{C}_{\mathbf{q}_j}\right)$.
$\hat{\mathbf{x}}_j$ is what ends up being transmitted by BS $j$. Due to the backhaul constraint, the mutual information $I(\mathbf{\hat{x}}_j, \mathbf{x}_j)$ must satisfy
\begin{align}
I(\mathbf{\hat{x}}_j, \mathbf{x}_j)  \le C_j. \label{eq:ConstraintCjQ}
\end{align}
Moreover, given the power constraints, the covariance of $\mathbf{\hat{x}}_j$, $\mathbf{C}_{\hat{\mathbf{x}}_j}$ must be such that
\begin{align}
\mathrm{Tr}\left[\mathbf{C}_{\hat{\mathbf{x}}_j}\right] &= \mathrm{Tr}\left[\mathbf{C}_{\mathbf{x}_j} +\mathbf{C}_{\mathbf{q}_j}\right] 
\le P_j. \label{eq:ub_sigmaqi}
\end{align}

The signal received at user $k$ is thus\footnote{Recall $\bar{k}$ is used to denote the 'other' user/base station.}
\begin{align}
y_k &= \sum_{j=1}^2 \mathbf{h}_{kj} \left[\mathbf{x}_j + \mathbf{q}_j\right]  + n_k
= \sum_{j=1}^2 \mathbf{h}_{kj} \mathbf{w}_{kj}s_k + z_k
\end{align}
where
\begin{align}
z_k = n_k + \sum_{j=1}^2 \mathbf{h}_{kj} \left[ \mathbf{w}_{\bar{k}j}s_{\bar{k}} + \mathbf{q}_j\right].
\end{align}

Let $\mathbf{w}_k = \left[\mathbf{w}_{k1}; \mathbf{w}_{k2}\right]$ be the joint precoding vector carrying user $k$'s symbols. $r_k$ will be upper bounded by
\begin{align}
&\log_2 \left(1+\frac{\left|\mathbf{h}_{k} \mathbf{w}_{k}\right|^2
}{
\sigma^2 + \left|\mathbf{h}_{k} \mathbf{w}_{\bar{k}}\right|^2
+ \sum_{j=1}^2 \mathbf{h}_{kj} \mathbf{C}_{\mathbf{q}_j} \mathbf{h}_{kj}^H
}\right) \label{eq:QR}
\end{align}

To tackle the design of the precoding and the quantization, we distinguish between the case where $N_t = 1$ and that when $N_t \ge 2$.
\subsubsection{$N_t = 1$}
In this case, the covariance matrix $\mathbf{C}_{\mathbf{q}_j}$ boils down to a single parameter, the quantization noise variance $\sigma^2_{\mathbf{q}_j}$, so that \eqref{eq:ConstraintCjQ}  becomes
\begin{align}
\log_2\left(1+\frac{|\mathbf{w}_{1j}|^2 + |\mathbf{w}_{2j}|^2}{\sigma^2_{\mathbf{q}_j}}\right) \le C_j. \label{eq:QRNt1}
\end{align}
Similarly, the power constraint at BS $j$ reduces to $|\mathbf{w}_{1j}|^2 + |\mathbf{w}_{2j}|^2 + \sigma^2_{\mathbf{q}_j} \le P_j$.

It is clear that the best rates require $\sigma^2_{\mathbf{q}_j}$ to be as small as possible. Thus, from the backhaul constraint, $\sigma^2_{\mathbf{q}_j} = \frac{|\mathbf{w}_{1j}|^2 + |\mathbf{w}_{2j}|^2}{2^{C_j}-1}$, and constraint \eqref{eq:QRNt1} can be transformed into a second order cone convex one. This is however not the case if $N_t \ge 2$, which is now treated separately below.

\subsubsection{$N_t \ge 2$}
$\mathbf{x}_j$'s covariance matrix, $\mathbf{w}_{1j}\mathbf{w}_{1j}^H + \mathbf{w}_{2j}\mathbf{w}_{2j}^H$, is rank 2.  Let $\mathbf{U}_j \boldsymbol{\Lambda}_j \mathbf{U}_j^H$ be its eigenvalue distribution, and let $\mathbf{U}_j^{(1)}$ be the two columns corresponding to the nonzero eigenvalues, referred to as $\lambda_{j,1}$ and $\lambda_{j,2}$. Premultiplying \eqref{eq:xQ} by $\mathbf{U}_j^{(1)~H}$ yields an equivalent channel
\begin{align}
\tilde{\hat{\mathbf{x}}}_j = \tilde{\mathbf{x}}_j + \tilde{\mathbf{q}}_j,
\end{align}
such that $\tilde{\hat{\mathbf{x}}}_j = \mathbf{U}_j^{(1)~H} \hat{\mathbf{x}}_j$, $ \tilde{\mathbf{x}}_j = \mathbf{U}_j^{(1)~H} \mathbf{x}_j$, $\tilde{\mathbf{q}}_j = \mathbf{U}_j^{(1)~H}\mathbf{q}_j$, all in $\mathbb{C}^2$.

$\mathbf{C}_{\tilde{\mathbf{x}}_j} = \textrm{diag}\left[\lambda_{j,1}, \lambda_{j,2}\right]$. \eqref{eq:ConstraintCjQ} becomes
\begin{align}
\sum_{i=1}^2 \log_2\left(1+\frac{\lambda_{j,i}}{\sigma^2_{\tilde{\mathbf{q}}_{j,i}}}\right) \le C_j.
\end{align}
whereas \eqref{eq:ub_sigmaqi} becomes $\sum_{i=1}^2 \left(\lambda_{j,i}+\sigma^2_{\tilde{\mathbf{q}}_{j,i}}\right) \le P_j$.
However, unlike the $N_t=1$ case, studying the above power and backhaul link constraints does not seem to offer a  simple characterization of the quantization noise covariance matrix, and attempts to solve the problem only guarantee a local optimum.

When determining the corresponding achievable rate region in Section \ref{sec:numres}, we use Matlab's \emph{fmincon} function.

\section{Numerical Results}\label{sec:numres}
Throughout the simulations, $C_1 = C_2$ and the common value is denoted $C$. Similarly, $P_1 = P_2$ and the common value is denoted $P$.
Since the rate region is established for one given channel instance, we illustrate the gains arising from finite shared messages over one example of a channel given by some arbitrary, yet fixed, coefficients. Later on we show Monte Carlo results obtained over fading channels.

\begin{figure}[htb]
\center{
\includegraphics[width=3in, height=3in, viewport = 100 230 500 542, clip]{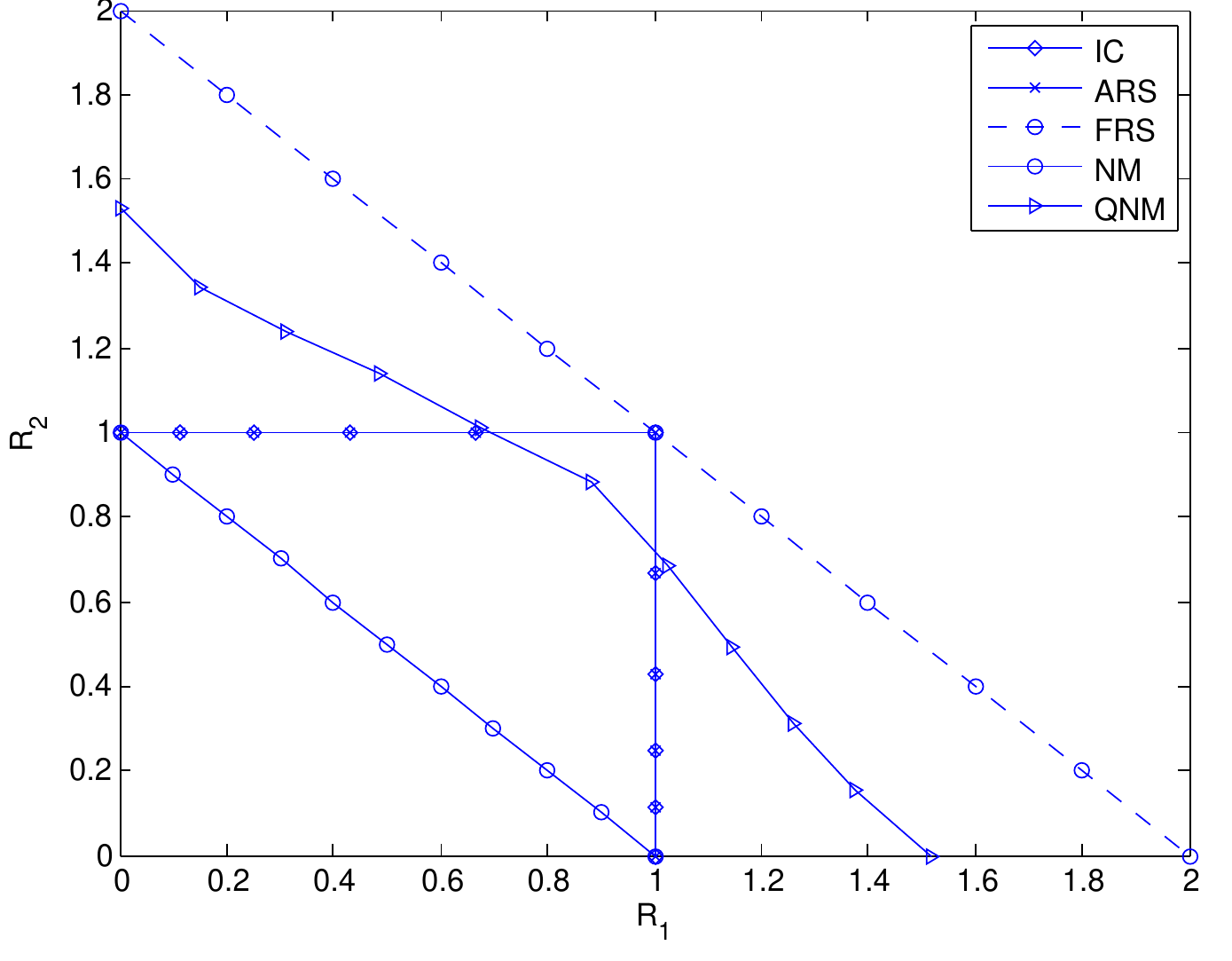}
\caption{Sample Rate Regions Comparison for 10dB SNR, $C = 1$ bits/sec/Hz}
\label{fig:N2SNR10C1}
}
\end{figure}

\begin{figure}[htb]
\center{
\includegraphics[width=3in, height=3in, viewport = 100 230 500 542, clip]{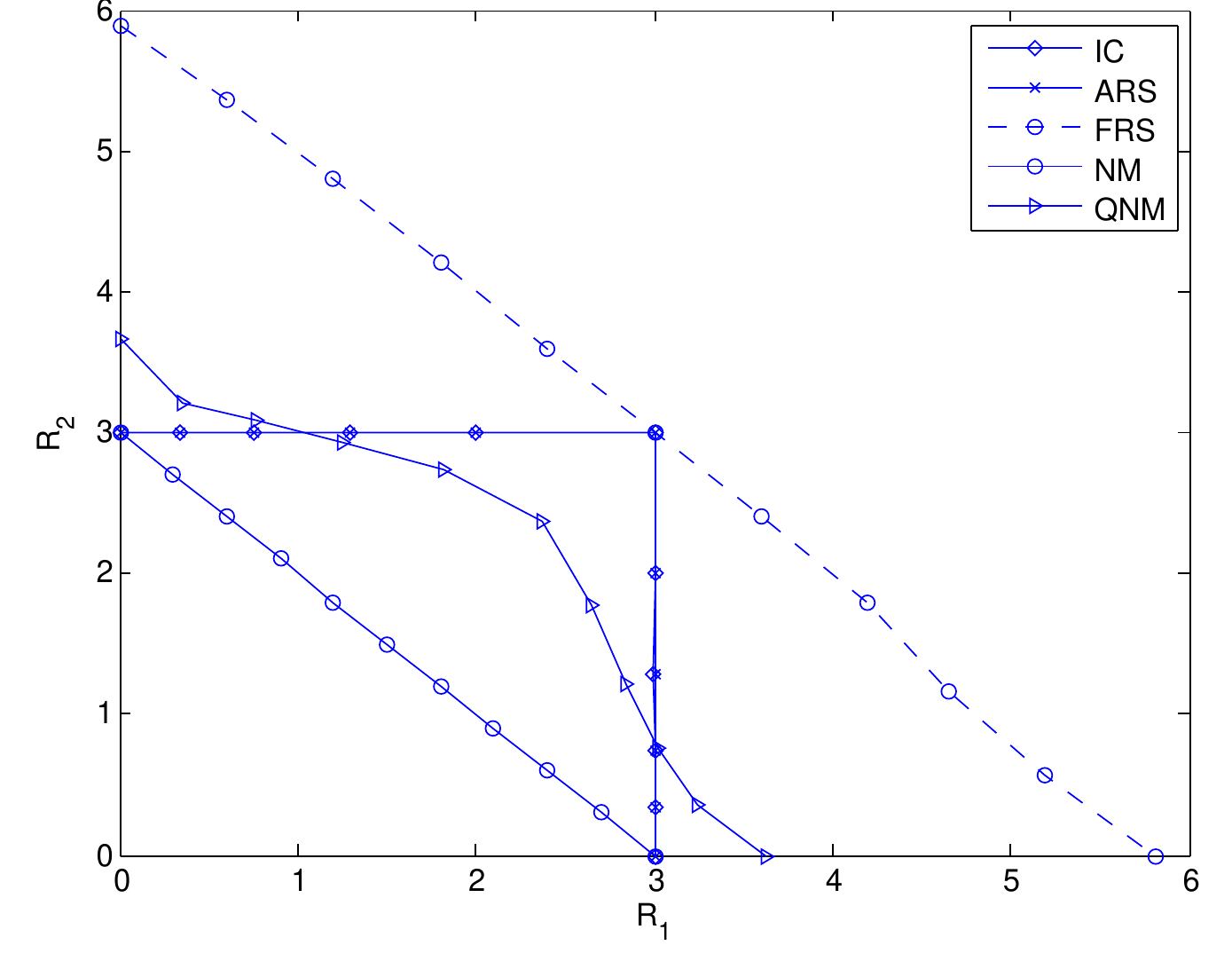}
\caption{Sample Rate Regions Comparison for 10dB SNR, $C = 3$ bits/sec/Hz}
\label{fig:N2SNR10C3}
}
\end{figure}

\begin{figure}[htb]
\center{
\includegraphics[width=3in, height=3in, viewport = 100 230 500 542, clip]{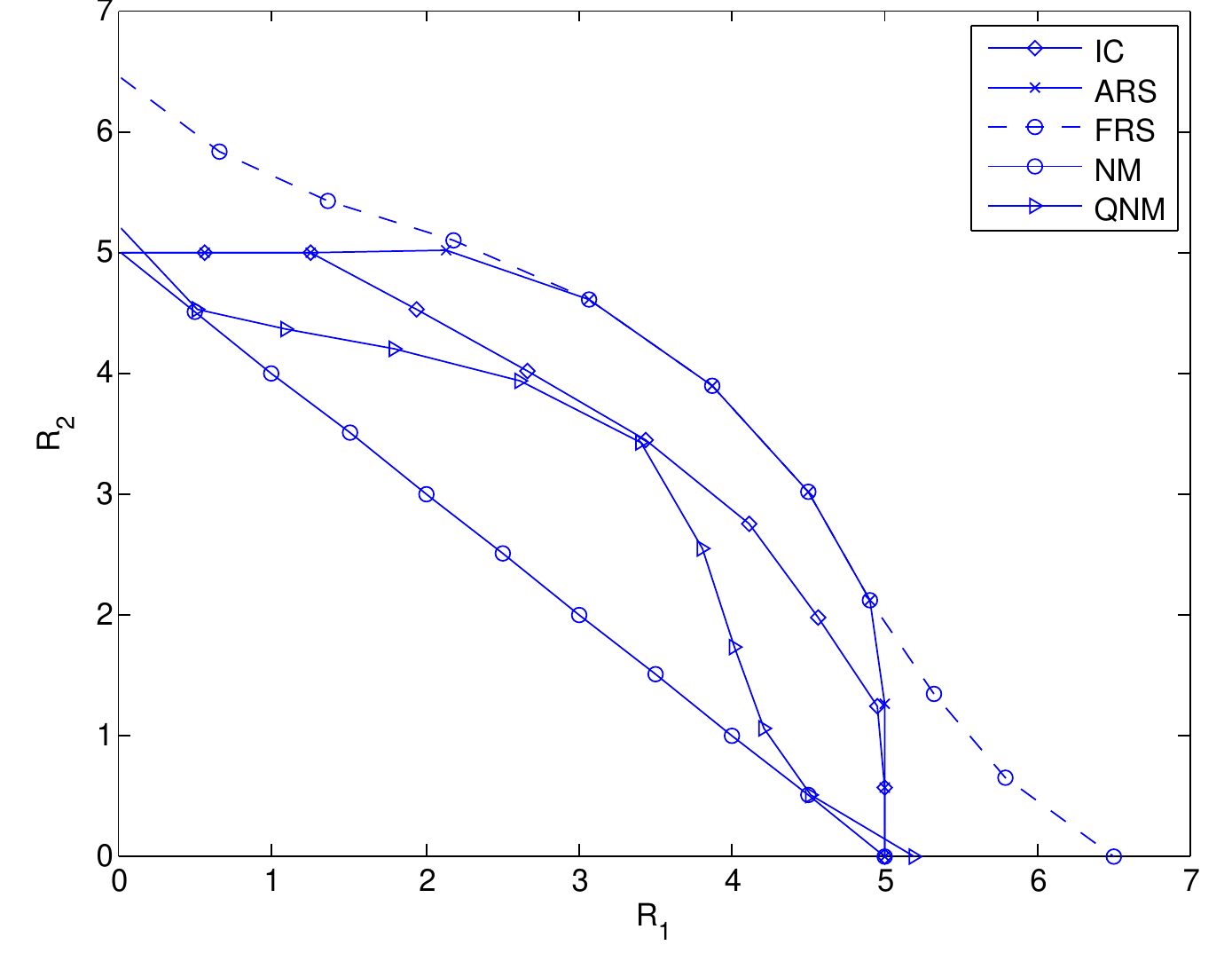}
\caption{Sample Rate Regions Comparison for 10dB SNR, $C = 5$ bits/sec/Hz}
\label{fig:N2SNR10C5}
}
\end{figure}

Figures \ref{fig:N2SNR10C1}-\ref{fig:N2SNR10C5} show, for the following channel
\begin{align}
\mathbf{h}_{11} &= [0.2939 - 1.1488\mathbf{i} ~  -1.5260 - 0.3861\mathbf{i}],\nonumber \\
 \mathbf{h}_{12} &= [  0.3963 - 0.2679\mathbf{i}  ~0.8306 + 0.6110\mathbf{i}],\nonumber \\
\mathbf{h}_{21} &= [  -0.7201 - 0.3025\mathbf{i} ~  -0.9658 - 0.1754\mathbf{i}],\nonumber \\
\mathbf{h}_{22} &= [0.1952 - 0.0026\mathbf{i} ~ 1.7096 + 0.4040\mathbf{i}],\nonumber
\end{align}
and different values of the backhaul constraints, the rate regions achieved for an SNR ($P/\sigma^2$) of 10 dB by the following different schemes:
\begin{itemize}
\item The proposed rate splitting scheme, which we label FRS (for Full Rate Splitting)
\item The rate splitting scheme studied in \cite{zakhour_izs10}, where private rates originate from only one of the two base stations ($r_{ij,p} = 0$, for $i \neq j$), which we label ARS (for Asymmetric Rate Splitting),
\item Beamforming on the interference channel ($r_{ii,p} = r_{i}, i = 1, 2$), labeled IC,
\item Network MIMO beamforming ($r_{i,c} = r_i, i = 1, 2$), labeled NM,
\item The quantized backhaul network MIMO scheme, labeled QNM.
\end{itemize}
As can be seen, depending on $C$, the FRS scheme may achieve a total sum rate of up to $2C$, which is the maximum possible: this is the case in Figure \ref{fig:N2SNR10C1} for example.
One can also note that if $C$ is relatively low, one may be better off giving up on a network MIMO approach, especially if the backhaul is used to forward the messages themselves.
As the backhaul capacity increases, the NM approach increases in appeal. The FRS and ARS approaches outperform it as $C$ increases until the point where both achieve the same rate region: when this happens, the system is no longer backhaul-limited and becomes limited by the achievable rate region over the air interface.

Note that simulations not presented here have shown that QNM might provide slightly better results over a portion of the rate region: Of course, the applicability of either scheme may be limited by the network infrastructure itself and where the network 'intelligence' is located.
Also, the rate region achieved by the FRS and ARS schemes are not always as smooth and can be nonconvex, similar to the QNM region, since we do not convexify the region by time sharing between different transmit strategies.

\begin{figure}[htb]
\center{
\includegraphics[width=3.5in, height=3in, viewport = 50 180 550 594, clip]{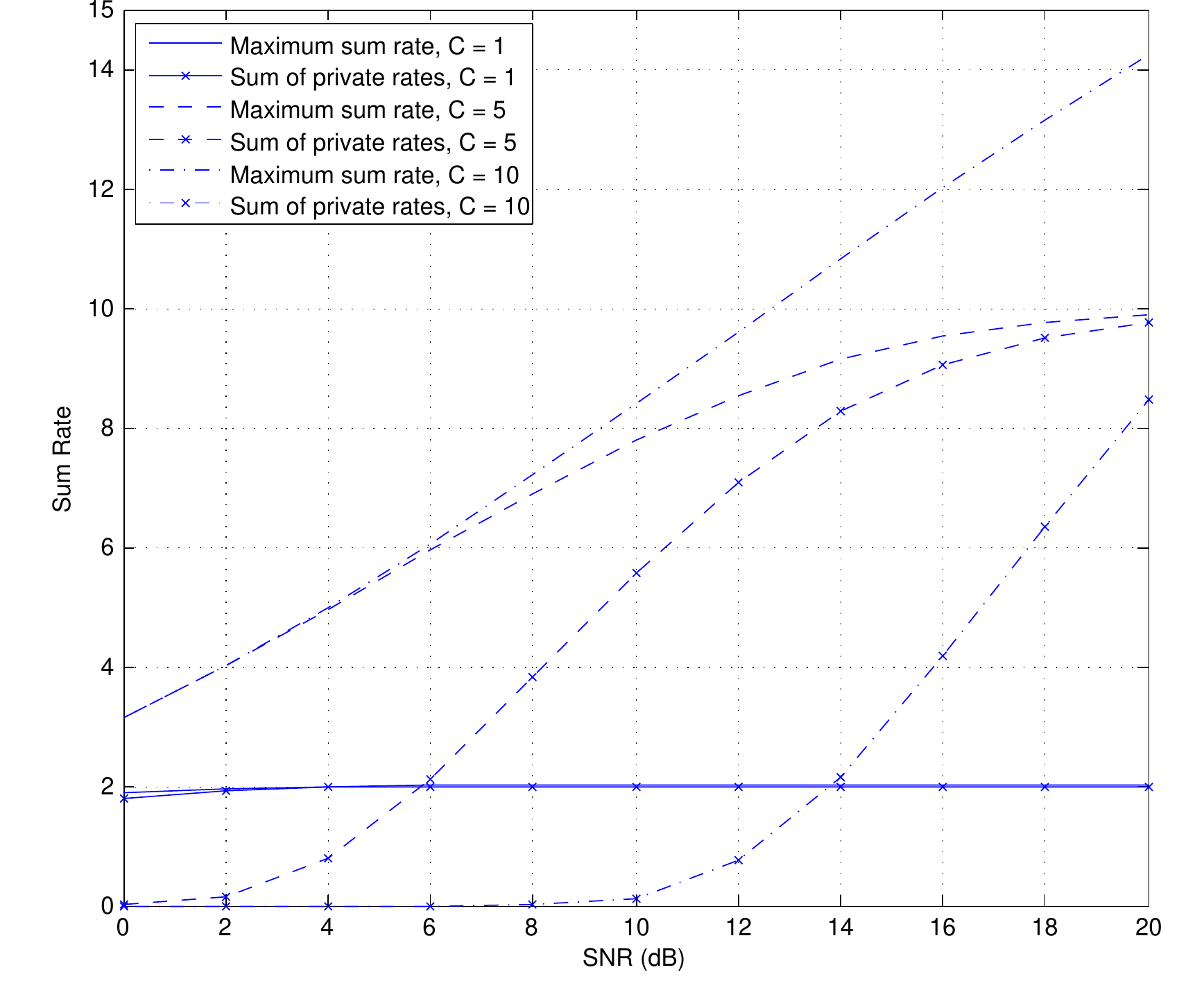}
\caption{Average maximum sum rate versus SNR for $C = 1, 5$ and 10 bits/sec/Hz, and and symmetric channels with cross channel variance .1. The figure also shows how much of the rates are in the form of private messages.}
\label{fig:epsilon01}
}
\end{figure}

\begin{figure}[htb]
\center{
\includegraphics[width=3.5in, height=3in, viewport = 50 180 550 594, clip]{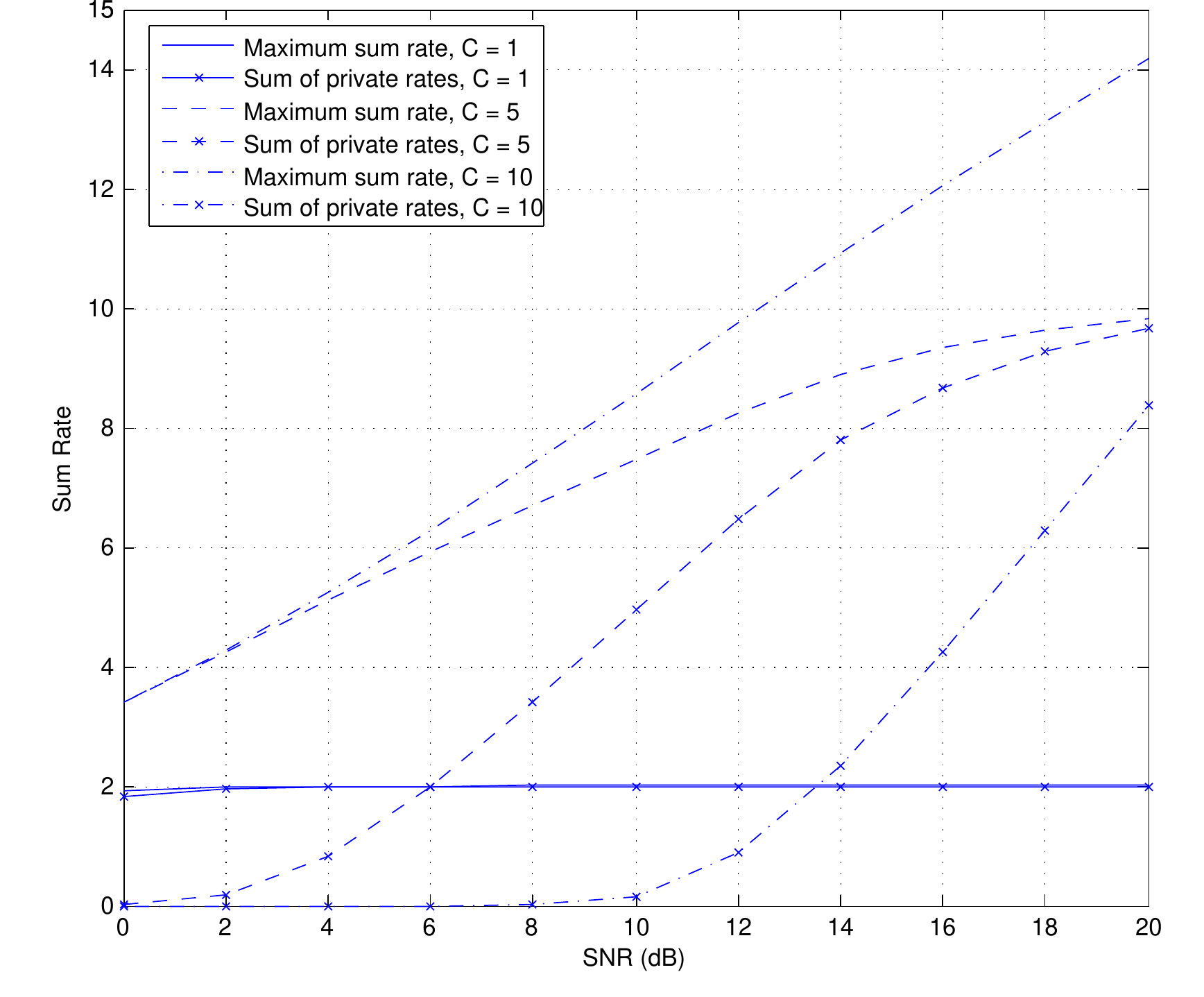}
\caption{Average maximum sum rate versus SNR for $C = 1, 5$ and 10 bits/sec/Hz, and and symmetric channels with cross channel variance .5. The figure also shows how much of the rates are in the form of private messages.}
\label{fig:epsilon05}
}
\end{figure}

Figures \ref{fig:epsilon01} and \ref{fig:epsilon05} show the maximum sum rates achieved by the FRS scheme averaged over 100 channel samples, for $N_t = 2$ and different values of the backhaul. The channels are assumed to be symmetric in the sense that
\begin{align}
\mathbf{h}_{ii} &\sim \mathcal{CN}\left(\boldsymbol{0}, \mathbf{I}\right), \quad i = 1,2\\
\mathbf{h}_{i\bar{i}} &\sim \mathcal{CN}\left(\boldsymbol{0}, \epsilon \mathbf{I}\right), i = 1, 2, \bar{i} = \mod(i, 2)+1
\end{align}
where the parameter $\epsilon$ quantifies the strength of the cross links (in an interference channel, this would be the strength of the interfering link).
For $C$ low, the maximum sum rate of $2C$ is achievable for quite low SNR. As $C$ increases the saturation of the sum rate at $2C$ occurs at higher SNR.
Also shown in the figures is how much of the total data rate comes from private messages. Our feasibility check as detailed in Section \ref{sec:FeasibleSR} does not seek to maximize the total private messages: it simply checks the corner points of the feasible rate region defined by the backhaul constraints for feasibility over the air interface and exits at the first instance of a feasible set of rates. However, for $C$ quite low, most of the data will be in the form of private messages, whereas as $C$ increases, private messages will be required for higher values of the SNR only. Thus, for $C = 10$, for an SNR lower than 10 dB, the sum rate can be maximized almost always by a network MIMO approach.

\section{Conclusion}
In this paper, we proposed to use the backhaul capacity to convey different types of messages: private messages transmitted from only one of the base stations, and shared messages jointly transmitted from several base stations. A corresponding achievable rate region for the two-cell setup was characterized and simulations have illustrated the benefit of the rate splitting approach adopted. The study shows how the portion of traffic that ought to be shared grows with the backhaul capacity limit.
Our rate splitting approach was compared to one relying on quantization, which it normally outperforms. Note that in both cases, better rates may be attained by dirty-paper coding, not considered here.

\appendices
\section{MAC with a common message}\label{app:mac_common}
For convenience, we reproduce the following result from \cite{kramer_07}, initially obtained by Slepian and Wolf \cite{slepian_Bell73}, where $I(.;.)$ denotes mutual information and $I(.;.|.)$ denotes conditional mutual information:
\begin{theorem}[MAC with a common message]\label{theo:mac} The sources put out statistically independent messages $W_0, W_1, W_2$ with $nR_0$, $nR_1$, $nR_2$ bits, respectively. The message $W_0$ is seen by both encoders and is called the \emph{common} message, whereas $W_1$ and $W_2$ appear only at the respective encoders 1 and 2, i.e. are \emph{private} to those encoders. Encoder 1 maps $(w_0, w_1)$ to a sequence $x_1^n \in \mathcal{X}_1^n$, encoder 2 maps $(w_0, w_2)$ to a sequence $x_2^n \in \mathcal{X}_2^n$, and the channel $P_{Y|X_1, X_2}(.)$ puts out a sequence $y^n \in \mathcal{Y}^n$. Consider a distribution $P_{UX_1X_2Y}$ that factors as $P_U P_{X_1|U} P_{X_2|U} P_{Y|X_1X_2}$. The following rate region, denoted $\mathcal{R}(P_U, P_{X_1|U}, P_{X_2|U})$, is achievable:
\begin{align}
R_1 < I(X_1; Y|X_2,U), \nonumber \\
R_2 < I(X_2; Y|X_1, U), \nonumber \\
R_1 + R_2 < I(X_1, X_2; Y|U), \nonumber \\
R_0 + R_1 + R_2 < I(X_1, X_2; Y). \label{eq:MAC_common}
\end{align}
One can further restrict attention to $|\mathcal{U}| \le \min \left(|\mathcal{Y}|+3, |\mathcal{X}_1||\mathcal{X}_2|+2\right)$.
The capacity of the thus defined MAC is the union of such regions,
\begin{align}
\mathcal{C}_{\mathrm{MAC}} = \bigcup_{P_U, P_{X_1|U}, P_{X_2|U}} \mathcal{R}(P_U, P_{X_1|U}, P_{X_2|U}).
\end{align}
\end{theorem}

In the setup considered, let $s_{i,c}$, $s_{ij,p}$, $j=1,2$, be  independent $\mathcal{CN}(0, 1)$ random variables, and define the transmit signal of BS $j$ to user $i$, $\mathbf{x}_{ij}$, as in \eqref{eq:xij}.
The received signal at user $i$, as given by \eqref{eq:y_i}, becomes
\begin{align}
y_i = \sum_{j = 1}^2 \mathbf{h}_{ij} \mathbf{x}_{ij} + \tilde{z}_i,
\end{align}
$\tilde{z}_i \sim \mathcal{CN}(0, \sigma^2_i)$ is the receiver noise plus interference, which from \eqref{eq:xij}, can be verified to be equal to
\begin{align}
\sigma^2_i &= \sigma^2 + \mathbb{E} \left|\sum_{j=1}^2 \mathbf{h}_{ij} \mathbf{x}_{\bar{i}j}\right|^2 \nonumber \\
&=  \sigma^2 + \sum_{j=1}^2 \left|\mathbf{h}_{ij}\mathbf{w}_{\bar{i}j,p}\right|^2 +\left|\mathbf{h}_{i}\mathbf{w}_{\bar{i},c}\right|^2.
\end{align}

In the proposed transmission scheme, $s_{i, c}$ is the equivalent of $U$ in \eqref{eq:MAC_common}.
Thus, the following rates are achievable, under perfect channel state information at the receiver (CSIR): 
\begin{align}
&r_{ij,p} < I(\mathbf{x}_{ij}; y_i | \mathbf{x}_{i\bar{j}}, \mathbf{w}_{i,c} s_{i, c}), \quad j = 1, 2
\label{eq:rijp}\\
&\sum_{j=1}^2 r_{ij,p} < I(\mathbf{x}_{i1}, \mathbf{x}_{i2}; y_i | \mathbf{w}_{i,c} s_{i, c}), \quad j = 1, 2
\label{eq:rip}\\
&r_i = \sum_{j=1}^2 r_{ij,p} + r_{i,c} < I(\mathbf{x}_{ii}, \mathbf{x}_{i\bar{i}}; y_i), \label{eq:ri}
\end{align}
One can easily verify that the mutual information expressions are the ones in \eqref{eq:MAC_rates}.

\section{}\label{app:Dual}
The equivalence of \eqref{eq:minPow} for fixed rates to a convex optimization is shown by
\begin{itemize}
\item Taking the positive semidefinite relaxation of the problem (see \cite{gershman_sp2010} for example), by introducing for each vector $\mathbf{w}$ a positive semidefinite matrix $\mathbf{V}$, which replaces $\mathbf{w}\mathbf{w}^H$;
\item Noting that the relaxed problem is convex and has zero duality gap;
\item Noting that the optimal matrices will be rank one.
\end{itemize}

Let
\begin{align}
&\Gamma_{1j,p} = 2^{r_{1j,p}}-1, \quad \Gamma_{2j,p} = 2^{c_j - r_{1j,p}}-1,  \nonumber \\
&\Gamma_{1,p} = 2^{\sum_j r_{1j,p}}-1, \quad \Gamma_{2,p} = 2^{\sum_j (c_j- r_{1j,p})}-1, \\
&\Gamma_i = 2^{r_i}-1, ~ i = 1,2,
\end{align}
and $\mathbf{R}_i = \mathbf{h}_i^H \mathbf{h}_i$, $\mathbf{R}_{ij} = \mathbf{h}_{ij}^H \mathbf{h}_{ij}$, the relaxation is given by
\begin{align}
\text{min. } & \sum_{i=1}^2 \left(\mathrm{Tr} \mathbf{V}_{i,c} + \sum_{j=1}^2 \mathrm{Tr} \mathbf{V}_{ij,p} \right)\\
&\Gamma_{ij,p} \left[\sigma^2 + \sum_{j'} \mathrm{Tr} \left[\mathbf{R}_{ij'}\mathbf{V}_{\bar{i}j',p}\right] +\mathbf{Tr}\left[\mathbf{R}_{i}\mathbf{V}_{\bar{i},c}\right]\right]\nonumber \\
& \le \mathrm{Tr} \left[\mathbf{R}_{ij}\mathbf{V}_{ij,p}\right], \quad  i, j = 1, 2\nonumber \\
&\Gamma_{i,p}\left[\sigma^2 + \sum_{j'} \mathrm{Tr} \left[\mathbf{R}_{ij'}\mathbf{R}_{\bar{i}j',p}\right] +\mathrm{Tr} \left[\mathbf{R}_{i}\mathbf{V}_{\bar{i},c}\right]\right] \nonumber \\
&\le \sum_{j=1}^2 \mathrm{Tr} \left[\mathbf{R}_{ij}\mathbf{V}_{ij,p}\right], i = 1, 2\nonumber \\
&\Gamma_i \left[\sigma^2 + \sum_{j'} \mathrm{Tr} \left[\mathbf{R}_{ij'}\mathbf{V}_{\bar{i}j',p}\right]  +\mathrm{Tr} \left[\mathbf{R}_{i}\mathbf{V}_{\bar{i},c}\right]\right] \nonumber \\
&\le \mathrm{Tr} \left[\mathbf{R}_{i}\mathbf{V}_{i,c}\right]+\sum_{j=1}^2 \mathrm{Tr} \left[\mathbf{R}_{ij}\mathbf{V}_{ij,p}\right], i = 1,2\nonumber\\
&\sum_{i=1}^2 \left(\mathrm{Tr}\left[\mathbf{D}_j \mathbf{V}_{i,c}\right] + \mathrm{Tr} \mathbf{V}_{ij,p}\right)  \le P_j, ~j = 1, 2. \nonumber
\end{align}

One can show that the optimal matrices will be rank one by considering individual useful and interfering terms corresponding to each one and noting that the maximization of the useful term subject to a power and an interference constraint will have a rank one optimal solution \cite{zhang_sp2010}.

Denote by $\lambda_{ij,p}$ the Lagrange coefficient associated with the constraint on the private rate to user $i$ from base station $j$; similarly, $\lambda_{i,p}$ is the Lagrange coefficient associated with the constraint on the sum of private rates to user $i$ and $\lambda_i$ is to Lagrange coefficient associated with the total rate constraint. Finally, let $\mu_j$  be the Lagrange coefficient associated with the power constraint at BS $j$.

For given rates, the dual problem \cite{boyd} of \eqref{eq:minPow}, and of its semidefinite relaxation, is
\begin{align}
\textrm{max. } & \sigma^2 \sum_i \left[\sum_{j} \lambda_{ij,p} + \lambda_{i,p} + \lambda_i \right] - \sum_{j} \mu_j P_j \nonumber \\
\textrm{s.t. } & \mathbf{I}_{2N_t} + \mu_1 \mathbf{D}_1 + \mu_2 \mathbf{D}_2+ \left(\sum_{j=1}^2 \lambda_{\bar{i}j,p} + \lambda_{\bar{i},p} +{\lambda}_{\bar{i}}  \right) \mathbf{h}_{\bar{i}}^H\mathbf{h}_{\bar{i}} \nonumber \\
& \ge \frac{\lambda_i}{\Gamma_{i}} \mathbf{h}_{{i}}^H\mathbf{h}_{{i}} \nonumber \\
&(1+ \mu_j)\mathbf{I}_{N_t} + \left(\sum_{j'=1}^2 \lambda_{\bar{i}j',p}  + \lambda_{\bar{i},p}  +{\lambda}_{\bar{i}} \right) \mathbf{h}_{\bar{i}j}^H\mathbf{h}_{\bar{i}j} \nonumber \\
&\ge \left(\frac{\lambda_{ij,p}}{\Gamma_{ij,p}} + \frac{\lambda_{i,p}}{\Gamma_{i,p}} + \frac{\lambda_i}{\Gamma_i}\right) \mathbf{h}_{ij}^H\mathbf{h}_{ij} \nonumber
\end{align}

Equivalently, and introducing the short-hand $s_i = \left(\sum_{j=1}^2 \lambda_{\bar{i}j,p} + \lambda_{\bar{i},p} +{\lambda}_{\bar{i}}  \right) $,
\begin{align}
\textrm{max. } & \sigma^2 \sum_i \left[\sum_{j} \lambda_{ij,p} + \lambda_{i,p} + \lambda_i \right] - \sum_{j} \mu_j P_j \nonumber \\
\textrm{s.t. } &\frac{1}{\mathbf{h}_{{i}}\left[\mathbf{I}_{2N_t} + \mu_1 \mathbf{D}_1 + \mu_2 \mathbf{D}_2+ s_i \mathbf{h}_{\bar{i}}^H\mathbf{h}_{\bar{i}}\right]^{-1}\mathbf{h}_{{i}}^H }  \nonumber \\
& \ge \frac{\lambda_i}{\Gamma_{i}} \label{eq:lambda_i} \\
&\frac{1}{\mathbf{h}_{ij}\left[(1+ \mu_j)\mathbf{I}_{N_t} + s_i \mathbf{h}_{\bar{i}j}^H\mathbf{h}_{\bar{i}j} \right]^{-1} \mathbf{h}_{ij}^H} \nonumber \\
&\ge \left(\frac{\lambda_{ij,p}}{\Gamma_{ij,p}} + \frac{\lambda_{i,p}}{\Gamma_{i,p}} + \frac{\lambda_i}{\Gamma_i}\right)  \label{eq:lambda_ip}
\end{align}

At the optimum, \begin{itemize}
\item \eqref{eq:lambda_i}  holds with equality.
\item  \eqref{eq:lambda_ip} becomes
\begin{align}
&\frac{1}{\mathbf{h}_{ij}\left[(1+ \mu_j)\mathbf{I}_{N_t} + s_i \mathbf{h}_{\bar{i}j}^H\mathbf{h}_{\bar{i}j} \right]^{-1} \mathbf{h}_{ij}^H}\nonumber \\
&-\frac{1}{\mathbf{h}_{{i}}\left[\mathbf{I}_{2N_t} + \mu_1 \mathbf{D}_1 + \mu_2 \mathbf{D}_2+ s_i \mathbf{h}_{\bar{i}}^H\mathbf{h}_{\bar{i}}\right]^{-1}\mathbf{h}_{{i}}^H }  \nonumber \\
&\ge \frac{\lambda_{ij,p}}{\Gamma_{ij,p}} + \frac{\lambda_{i,p}}{\Gamma_{i,p}}
\end{align}
Let $j_{i, \min}$ denote the index of the user for which the left-hand side of the above equation is the smallest, and $j_{i, \max}$ the user for which it is the largest; let $t_{i, j_{\min}}$, $t_{i, j_{\max}}$ denote these values, respectively. Then
\begin{align}
&\lambda_{i,p} = t_{i, j_{\min}} \Gamma_{i,p} \\
&\lambda_{ij_{i, \max},p} = \Gamma_{ij_{i, \max},p} \left(t_{i,j_{\max}} - t_{j_{i,\min}}\right) \\
&\lambda_{ij_{i,\min},p} = 0
\end{align}
\end{itemize}

From the KKT conditions, this implies that constraints associated with $\Gamma_i$ and $\Gamma_{i,p}$ will hold with equality, as will  those associated with $\Gamma_{ij_{i, \max}, p}$, which is to be expected from what is known about the MAC channel \cite{cover}.

Note that if the $j_{i, \max}$ were known beforehand, the constraints associated with $j_{i, \min}$ (they will be guaranteed to hold at the solution) can be ignored and the resulting problem can be shown to be convex in a similar manner to the problem treated in \cite{zakhour_izs10}  (using results from \cite{yu_sp07} among others).

\bibliographystyle{IEEEtran}
\bibliography{zakhour_biblio}
\end{document}